\documentclass[twocolumn,amsmath,amssymb,pra,a4paper]{revtex4}
\usepackage{graphicx}
\usepackage{float}
\usepackage{epsfig}
\usepackage{subfigure}
\usepackage{pstricks}
\usepackage{color}

\newcommand{\Tr}{\mathop\mathrm{Tr}}

\newcommand{\ket}[1]{\left|#1\right>}
\newcommand{\bra}[1]{\left<#1\right|}

\date{\today}
\begin{document}
\title{Entanglement generation by interaction with semiclassical radiation}
\author{Amir Leshem and Omri Gat}
\affiliation{Racah Institute  of  Physics, Hebrew University
of Jerusalem, Jerusalem 91904, Israel}
\begin{abstract}
We address a fundamental issue in quantum mechanics and quantum information theory, the generation of an entangled pair of qubits that interact solely through a third, semiclassical degree of freedom, in the framework of cavity quantum electrodynamics. We show that finite, though not maximal, entanglement is obtainable in the classical limit, at the price of a diverging effective interaction time. The optimal atomic entanglement derives from a trade-off between the atomic entanglement in a sub-wave packet and the purity of the atomic state. Decoherence by photon loss sets an upper limit on the degree of excitation of the cavity mode, beyond which the achievable entanglement decreases as the inverse mean photon number to the sixth power.
\end{abstract}
\maketitle
\section{Introduction}
Field-atom entanglement is one of the hallmarks of strongly interacting cavity quantum electrodynamics (CQED) systems. This fundamental process is often used as a building block in the formation of entangled states of two or more atoms that serve as a resource for quantum information processing. A notable example is the generation of a two-atom Bell state by means of consecutive interaction with a resonant microwave cavity mode prepared in the vacuum state \cite{ens1}. In effect, the cavity mode stores the entanglement with one atom, that is then transferred to the second atom. The question addressed in this work is: Can entanglement be generated when the initial field state becomes semiclassical? On one hand quantum fluctuations persist also in the semiclassical limit, while on the other hand the correspondence principle states that in this limit classical physics should be recovered, where the notion of entanglement is not even defined.

The semiclassical states of the field mode are wave packets that are localized in phase space, where the canonical coordinates are the field quadratures, such as coherent states and squeezed coherent states, and have a large mean photon number. When a two-level atom interacts with a coherent state wave packet, its polarization undergoes Rabi oscillations whose amplitude exhibits collapse-revival dynamics \cite{eberly} as a consequence of the splitting of the initial wave packet in two mutually orthogonal sub-wave packets, and of their periodic collisions \cite{geaprl,geapra}. Each sub wave-packet is a product of squeezed field state and an atomic state. After splitting, the atom-field system is highly entangled, but slow atomic state evolution turns the field-atom state again into a product state \cite{geaprl,geapra}, and the field state is a superposition of two well-separated wave packets, sometimes called a Schr\"odinger cat state. This semiclassical dynamics can be described as a flow in a double phase space \cite{jc}.

\begin{figure*}[htb]
\includegraphics[width=0.4\linewidth]{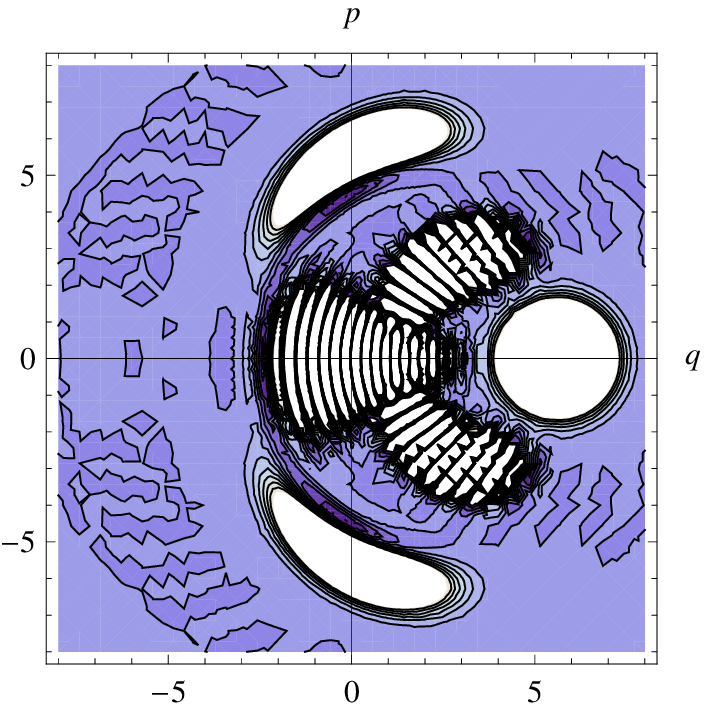}\hspace{1cm} \includegraphics[width=0.4\linewidth]{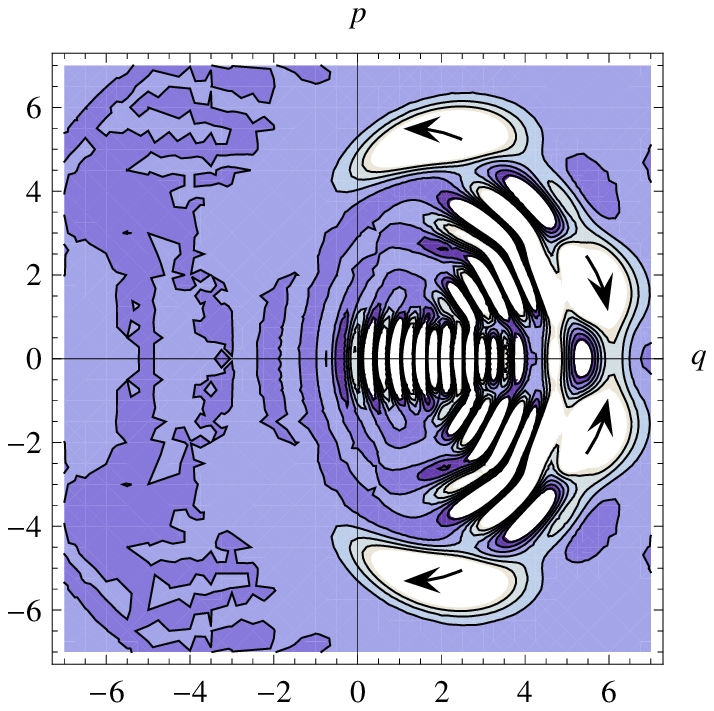}
\caption{\label{fig:field}The Wigner function of the field state $\Tr_\text{atoms}\ket{2}\bra{2}$, for unequal effective interaction times $\gamma_1\ne\gamma_2$ (left) and equal interaction times (right), numerically calculated for $\alpha=4$. In the left panel the arrows signify the direction of phase space rotation of the four sub wave packets. The Wigner function of the initial coherent state wave packet is circularly symmetric, and the deformation of the four sub wave packets in the left panel, and the two flank sub-wave packets in the right panel is a consequence of the squeezing.
}
\end{figure*}

Here we study a field mode interacting consecutively with two two-level atoms. The interaction of the field wave packet with the first atom causes it to split, as explained, and to become entangled with the atom. When the interaction with the second atom begins, each of these sub-wave packets splits again, and the atom-field-atom system becomes a superposition of four wave packets (see Sec. \ref{sec:afa}, Fig.\ \ref{fig:field}). \emph{Atom-atom} entanglement can only be obtained if two or more of the sub-wavepackets overlap, and this occurs when the normalized interaction times of the field with the two atoms are equal. The system state is then in general a superposition of three wave packets, and the atom-atom state is an entangled mixed state. 

We show in Sec.\ \ref{sec:aa} that it is possible to generate entanglement that reaches a finite, less than maximal, limit when the photon number tends to infinity, although the time required to generate the entangled state diverges in this limit. The optimal interaction time is determined as a trade-off between the mixing of three pure atomic states and the degree of entanglement of one of them, as shown in Fig.\ \ref{fig:2atom}. We consider the effects of photon loss in Sec.\ \ref{sec:dec} and show that the entanglement generation is degraded by dephasing of the sub wave packets, and that a coherent superposition of wave packets cannot be maintained when the mean photon number is larger than the inverse decoherence rate to the $\frac23$ power. Beyond this point entanglement generation proceeds only through small quantum fluctuation and it decreases as the inverse of the mean photon number to the sixth power, see Fig.\ \ref{fig:decoherence}.

The analysis demonstrates that a semiclassical preparation can serve as mediator for appreciable entanglement of atoms, if the decoherence is weak enough. For this purpose it is  necessary that the field becomes entangled with the atoms and evolve into macroscopic superposition states. In contrast, when decoherence is too strong to allow wave packet splitting, atom-atom entanglement is generated only through small quantum fluctuations, and it is therefore weak.

The resonant interaction of a field mode in excited states with two atoms has been investigated theoretically in \cite{mondragon,tcm,tcm2} for simultaneous interaction and \cite{nayak1,nayak2} for consecutive interaction, revealing wavepacket splitting, slow atomic state evolution, and collapse-revival dynamics of entanglement similar to those observed in one atom-field system. 
The off-resonant interaction of a field wave packet with two atoms was studied experimentally and theoretically in \cite{ens-dec,bppaper} as a means of measurement of the cavity decay rate and its effect on the dephasing of Schr\"odinger cat states. The steady state entanglement of two atoms in a highly excited leaky cavity was studied \cite{qd}. Our focus is the process of entanglement of two two-level systems by interaction with an intermediary that tends to the semiclassical limit, for which CQED provides a prominent realization.

Field-mediated atom-atom entanglement is often carried out in the highly-detuned Raman regime \cite{guo,ens2} where the field is only virtually excited to avoid field-induced decoherence. In a sense this is the opposite limit to the one studied here, and indeed we show that the atom-atom entanglement is very sensitive to dephasing of the field wave packet. Furthermore, although it is often argued that in the Raman regime entanglement is independent of the field state, the Raman regime entails an arbitrarily large detuning when the field state photon number tends to infinity.

\section{Atom-field-atom interaction}\label{sec:afa}

We study the dynamics of two two-level atoms interacting consecutively on resonance with a single electromagnetic mode. Modelling the interaction by the Jaynes-Cummings Hamiltonian in the rotating wave approximation, the Hamiltonian is $H(t)=H_1,\,0<t<t_1$, and $H_2,\,t_1<t<t_1+t_2$, where
\begin{equation}\label{eq:h}
H_k=\hbar\omega(a^\dagger a+\sigma_k^\dagger\sigma_k)+\hbar \Omega_k(a^\dagger\sigma_k+a\sigma_k^\dagger)
\end{equation}
where $\omega$ is the frequency of the electromagnetic mode with energy states $\ket{n}$, $n=0,1,\ldots~$ created by $a^\dagger$, and coherent states $\ket{\alpha}$, $\hbar\omega$ is the level spacing of the two-level atoms with energy states $\ket{g}_k,\ket{e}_k$ (atom k) raised by $\sigma^\dagger_k$, and $\hbar \Omega_k$ are the 
field-atom interaction energies. The system is prepared in a product state $\ket{\alpha}\otimes\ket{g}_1\otimes\ket{g}_2$.

When the field interacts with the first atom, the energy states are polariton states $\ket{n}_{\pm}\equiv\frac{1}{\sqrt2}(\ket{n-1}\otimes\ket{e}_1\pm\ket{n}\otimes\ket{g}_1)$, with energies $\hbar(n\omega\pm\sqrt{n}\Omega_1)$, and the absolute ground state $\ket{0}\otimes\ket{g}_1$. The field-atom state evolves into a superposition of two products of a squeezed coherent state and an atomic state \cite{eberly,geaprl,geapra}. Expressing the Hamiltonian in terms of polariton number operator $\hat n$, and the projections $P_\pm$ on the $\pm$ subspaces \cite{jc},
$H_1=H_+P_++H_-P_-$, where $H_\pm=\hbar(\hat n\omega\pm \sqrt{\hat n}\Omega_1)$,
the initial state is $\frac1{\sqrt2}(\ket{\alpha}_+-\ket{\alpha}_-)$, where 
$\ket{\alpha}_\pm=\pm\sqrt{2}P_\pm(\ket{\alpha}\otimes\ket{g}_1)$, and the field-first atom state at time $t_1$ is
\begin{equation}
\ket{1}=\ket{1_+}-\ket{1_-}\ ,\quad \ket{1_\pm}={\textstyle\frac1{\sqrt2}}e^{-\frac{i}{\hbar}H_\pm t_1}\ket{\alpha}_\pm
\end{equation}

For large $|\alpha|$ the dynamics generated by $H_\pm$ is well-approximated by classical dynamics in the polariton phase spaces, that is, phase spaces corresponding to the $\pm$ sub-Hilbert spaces, where the canonical coordinates $q,p$ are the field quadratures. For convenience, we join a $\ket{0}_\pm$ state to the $\pm$ subspaces (respectively) with occupations that remain exponentially small throughout. The initial states in both phase spaces are then coherent state Gaussian wave packets, and they evolve according to the rules of semiclassical phase space dynamics \cite{lj86}. In this approximation the wave packet evolution is determined completely by classical data generated by the classical Hamiltonians 
\begin{equation}
H_\pm^{\text{(cl)}}=\textstyle\frac{1}{2}(q^2+p^2+\hbar) \omega +\sqrt{\frac{\hbar}{2}(q^2+p^2+\hbar)}\Omega_1
\end{equation}
obtained from the Weyl phase-space representation of the quantum Hamiltonians. $H_\pm^{\text{(cl)}}$ generate nonlinear oscillations, that is rotation in phase space with an amplitude-dependent frequency.

The wave packet propagation consists of three parts: A phase-space translation from the initial point $(q,p)$, $\alpha=\frac{1}{\sqrt2}(q+ip)$ to the final point $(q_\pm,p_\pm)$, $\alpha_\pm=\frac{1}{\sqrt2}(q_\pm+ip_\pm)$ determined by the classical trajectory of the center of the wave packet, squeezing determined by the phase-space deformation generated by the nonlinear oscillations with squeeze parameters $\xi_\pm$, and an overall phase factor $e^{-i\phi_\pm}$ determined by the classical action of the classical orbits, so that the wave packets evolve to the squeezed coherent state
\begin{align}
\ket{1_\pm}&=e^{-i\phi_\pm}\ket{\alpha_\pm,\xi_\pm}\label{eq:af-pol}
\end{align}
where 
\begin{equation}\ket{\alpha,\xi}_\pm=e^{\alpha b^\dagger-\alpha^* b} e^{\frac12(\xi^*b^2-\xi(b^\dagger)^2)}\ket{0}_{\pm}
\end{equation}
$b$ being the polariton annihilation operator defined with the usual properties $b\ket{n}_{\pm}=\sqrt{n}\ket{n-1}_\pm$, $[b,b^\dagger]=1$.
The values of the wave packet parameters in a frame rotating with angular speed $\omega$
are $\phi_\pm=\phi(\pm\gamma_1)$, $\alpha_\pm=\alpha(\pm\gamma_1)$, $\xi_\pm=\xi(\pm\gamma_1)$ with
\begin{align}
\phi(\gamma)&=\textstyle\frac{\gamma}{2}|\alpha|^2-\frac12\arctan(\frac14\gamma)\\
\alpha(\gamma)&=\alpha e^{-\frac{i}{2}\gamma}\\
\xi(\gamma)&= \textstyle\text{arcsinh}(\frac14\gamma)e^{i(\gamma+\text{arccot}(\frac14\gamma))}
\end{align}
Here $\gamma_1=\frac{\Omega_1t_1}{|\alpha|}$ is twice the phase space angle of rotation, clockwise and counter-clockwise (respectively), of the $+$ and $-$ wave packet; $\frac{\Omega_1}{2|\alpha|}$ is the classical (angular) frequency of nonlinear oscillations.

The atom-field states can be expressed in terms of \emph{photon} states with the help of the identity 
$\ket{n}_\pm=\frac1{\sqrt2}(a\frac{1}{\sqrt{a^\dagger a}}\ket{n}\otimes\ket{e}_1\pm\ket{n}\otimes\ket{g}_1)$ as
\begin{align}\label{eq:af-ph}
\ket{1_\pm}&=\textstyle e^{-i\phi(\pm\gamma_1)}\ket{\alpha(\pm\gamma_1),\xi(\pm\gamma_1)}\nonumber\\&\otimes(e^{i(\gamma_0\mp\frac12\gamma_1)}\ket{e}_1\pm\ket{g}_1)
\end{align}
where $\gamma_0=\arg\alpha$. Here we used the fact that the squeezed state wavepackets are localized in phase-space, so that they are approximate eigenstates of $a$ and $a^\dagger$ with eigenvalues $\alpha(\pm\gamma_1)$ and $\alpha(\pm\gamma_1)^*$ (respectively).

The atom-field state $\ket{1}$ is therefore a superposition of two wavepackets that are well-separated in phase space except when $\gamma_1$ is close to an integer multiple of $2\pi$. The splitting of the wave packet generates field-atom entanglement that is almost maximal when $\gamma_1\gtrsim|\alpha|^{-1}$ and decreases to zero for $\gamma_1=\pi$ \cite{geaprl}.


The interaction with the second atom proceeds analogously. The final field-two atom state at time $t_1+t_2$ is
\begin{equation}\textstyle
\ket{2}=e^{-\frac{i}{\hbar}H_2t_2}(\ket{1_+}-\ket{1_-})\otimes\ket{g}_2
\end{equation}
where each of the two wave packets $\ket{1}_\pm\otimes\ket{g}_2$ splits again to two sub wavepackets, that continue to undergo phase-space rotation and squeezing according to the field-second atom polariton sign. $\ket{2}$  is therefore a superposition of \emph{four} localized wave packets labeled by two sign choices
\begin{align}\textstyle
&\ket{2}=\frac12(\ket{2_{++}}-\ket{2_{+-}}-\ket{2_{-+}}+\ket{2_{--}})\\
&\ket{2_{rs}}=\textstyle\frac12e^{-i\phi(\Gamma_{rs})}\label{eq:afa-ph}
\ket{\alpha(\Gamma_{rs}), \xi(\Gamma_{rs})}\\
&\quad\otimes(e^{i(\gamma_0-\frac12r\gamma_1)}\ket{e}_1+r\ket{g}_1)\otimes(e^{i(\gamma_0-\frac12\Gamma_{rs})}\ket{e}_2+s\ket{g}_2) \nonumber
\end{align}
where $\Gamma_{rs}=(r\gamma_1+s\gamma_2)$ and $\gamma_2=\frac{\Omega_2t_2}{|\alpha|}$.

\section{Atom-atom entanglement}\label{sec:aa}
\begin{figure*}[htb]
\includegraphics[width=0.45\linewidth]{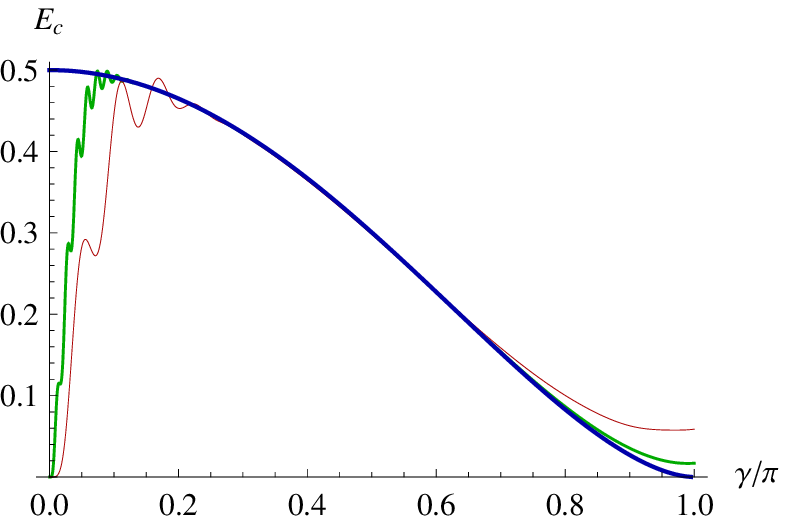} \includegraphics[width=0.45\linewidth]{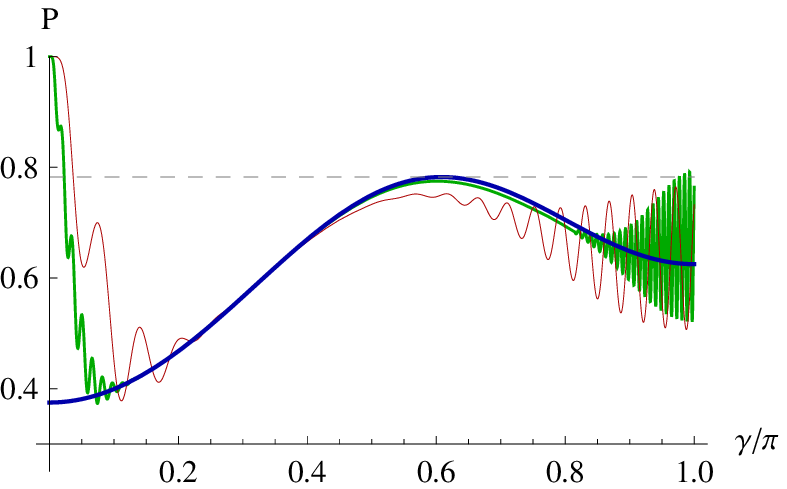}
 \caption{\label{fig:mixing}The entanglement entropy $E_c$ of the center wavepacket two-atom state $\ket{c}$ (left), and the purity $P$ of the full two-atom state $\rho_a$  (right)  as a function of the phase-space rotation angle $\gamma$, shown for $\alpha=4$ (thin red), $\alpha=8$ (medium green), and asymptotically for $|\alpha|\to\infty$ (thick blue). The initial rise and later drop in the degree of entanglement of the full two-atom state can be understood as a trade-off between the increase in $P$ and the decrease of $E_c$. $P(\alpha\to\infty)$ is discontinuous at $\gamma=0,\pi$, and these points are excluded from the graph.}
\end{figure*}
\begin{figure*}[htb]
\includegraphics[width=0.45\linewidth]{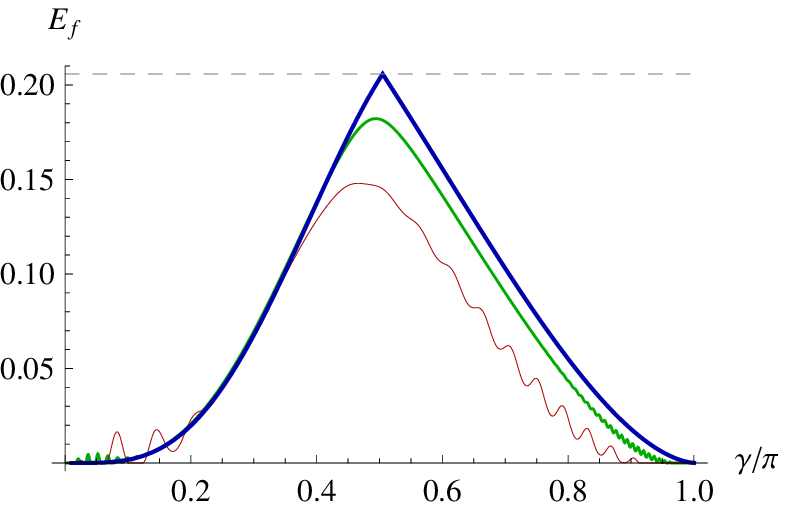}\hspace{8mm}\includegraphics[width=0.45\linewidth]{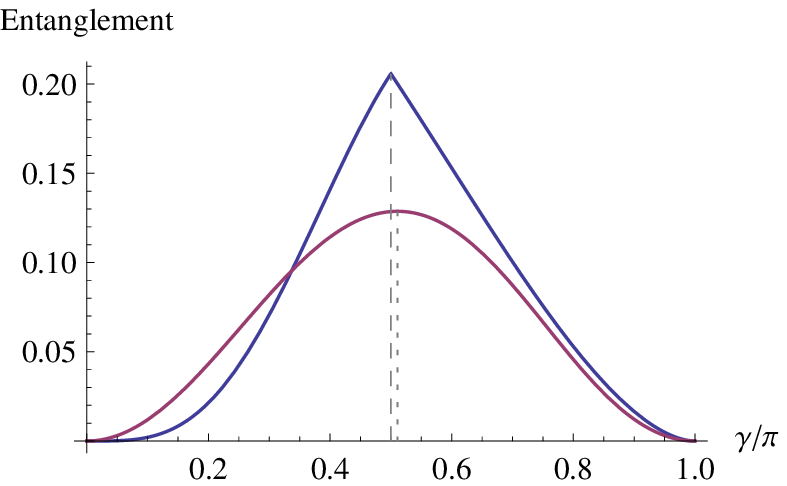}
 \caption{\label{fig:2atom}Left: The entanglement of formation $E_f$ of the two-atom state $\rho_a$ as a function of the phase-space rotation angle $\gamma$, shown for $\alpha=4$ (thin red), $\alpha=8$ (medium green), and asymptotically for $|\alpha|\to\infty$ (thick blue). Right: The entanglement of formation (upper, blue) and negativity (lower, violet) as a function of $\gamma$ for $|\alpha|\to\infty$. The two entanglement monotones attain their maxima (shown by vertical dashed lines) at close but different values of $\gamma$, showing that they are nonequivalent.}
\end{figure*}

The field-two atom state $\ket{2}$ is a superposition of four sub-wave packets each with a well-defined atomic product state. Superpositions of product states can give rise to entanglement, but for most values of $\gamma_1$ and $\gamma_2$ the four sub-wave packets are separate in phase space (see Fig.\ \ref{fig:field}), and label the different atomic components, so that the two-atom density matrix $\rho_a=\Tr_\text{field}\ket{2}\bra{2}$ is a classical mixture of four product states in the limit $|\alpha|\gg1$, which is by definition separable. The atomic states form a coherent superposition when the corresponding sub-wave packets overlap in phase space.
We therefore let $\gamma_1=\gamma_2=\gamma$ so that the field factors in the $\ket{2_{+-}}$ and $\ket{2_{-+}}$ terms are identical and equal to the initial coherent state $\ket{\alpha}$, while being separate from the field factors of $\ket{2_{++}}$ and $\ket{2_{--}}$, unless $\gamma$ is an integer multiple of $2\pi$. If $\gamma$ is not an integer multiple of $\pi$, the field factors of $\ket{2_{++}}$ and $\ket{2_{--}}$ are also phase-space separate, so that the atomic state is a mixture
\begin{equation}\textstyle\label{eq:densitya}
\rho_a=\frac14(\ket{l}\bra{l}+2\ket{c}\bra{c}+\ket{r}\bra{r})
\end{equation} 
of the states
\begin{align}
\ket{l}=&\textstyle\frac12(\ket{g}_1- e^{i\gamma_0}e^{\frac{i}{2}\gamma}\ket{e}_1)\otimes(\ket{g}_2- e^{i\gamma_0}e^{i\gamma}\ket{e}_2)\ ,\notag\\
\ket{c}=&\textstyle\frac1{\sqrt2}(e^{2i\gamma_0}\cos(\frac12\gamma)\ket{e}_1\otimes\ket{e}_2\notag\\
&\textstyle+ie^{i\gamma_0}\sin(\frac12\gamma)\ket{e}_1\otimes\ket{g}_2-\ket{g}_1\otimes\ket{g}_2)\ ,\ \text{and}\notag\\
\ket{r}=&\textstyle\frac12(\ket{g}_1+ e^{i\gamma_0}e^{-\frac{i}{2}\gamma}\ket{e}_1)\otimes(\ket{g}_2+ e^{i\gamma_0}e^{- i\gamma}\ket{e}_2)\ .\notag
\end{align}

The state $\ket{c}$ is entangled for all $\gamma$ not an integer multiple of $\pi$, with monotonically decreasing entanglement entropy as a function of $\gamma$ between $0$ and $\pi$ in the limit $|\alpha|\to\infty$ (see Fig.\ \ref{fig:mixing}), while the states $\ket{l}$ and $\ket{r}$ are separable. The full two-atom state is entangled if $\rho_a$ cannot be expressed as a mixture of separable states. The entanglement of formation \cite{wootters} is an entanglement measure that vanishes if and only if the state is separable, displayed in Fig.\ \ref{fig:2atom} that shows that the two atoms are entangled for most values of $\gamma$ for finite $\alpha$, and for all $\gamma$ except integer multiples of $\pi$ in the limit $|\alpha|\to\infty$. When nonzero, the entanglement of formation of the two-atom state, being equal to the minimal weighted average entanglement entropy of pure states mixed in $\rho_a$, obtains its maximum of $\approx0.21$ at a kink singularity at $\gamma=\frac\pi2$. Unfortunately, there exist nonequivalent entanglement monotones for mixed states \cite{hor}. As an example, the negativity, that is minus the sum of the negative eigenvalues of the partial transpose of $\rho_a$, obtains its maximum at $\gamma \approx0.511\pi$ (see Fig.\ \ref{fig:2atom}). This result indicates that although there is no precise meaning for optimal interaction time, values of $\gamma$ near $\frac12\pi$ are best for entanglement generation in this system. Furthermore, changing the initial condition of the second atom to $\ket{e}$ leaves the entanglement of formation unchanged, and slightly shifts the maximum of the negativity to  $\gamma\approx0.513\pi$.

An important aspect of the two-atom entanglement generation process is that although the atom-field entanglement, and therefore the entanglement entropy of the center wave packet $\ket{c}$, reaches maximal value after a short effective interaction time $\gamma$ of $O(|\alpha|^{-1})$, the degree of entanglement of the full two-atom state is very low for such small $\gamma$. This is a result of the incoherent mixing of $\ket{c}$ with the flank wave packets $\ket{l},\ket{r}$ whose overlap with $\ket{c}$ is low for small $\gamma$, as follows from the low purity of $\rho_a$ for these $\gamma$, see Fig.\ \ref{fig:mixing}.  When $\gamma$ increases, the entanglement entropy of $\ket{c}$ decreases until it becomes separable for $\gamma=\pi$, but at the same time the overlap of $\ket{c}$ with $\ket{l}$ and $\ket{r}$, and consequently the purity of $\rho_a$, increase. As a result, the degree of entanglement of $\rho_a$ increases initially, reaches a maximum near $\gamma=\frac12\pi$, and decreases to 0 for $\gamma=\pi$, as shown in Fig.\ \ref{fig:2atom}.
Although Eq.\ (\ref{eq:densitya}) is not valid for $\gamma$ an integer multiple of $\pi$, the preceding argument is valid for these values of $\gamma$ showing that the entanglement of formation is a continuous function of $\gamma$.


\section{Effects of decoherence}\label{sec:dec}
\begin{figure*}[htb]
\subfigure{\epsfig{file=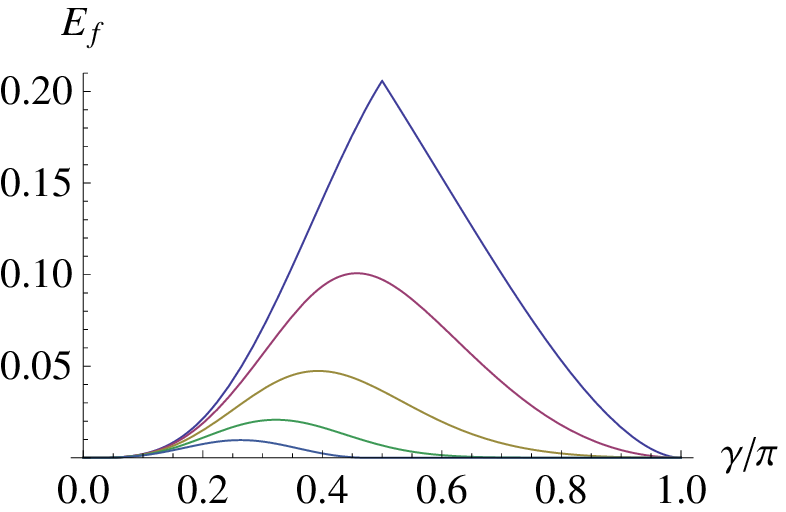,width=0.45\linewidth}}\hfill
	\subfigure{\hspace{-0.2cm}\epsfig{file=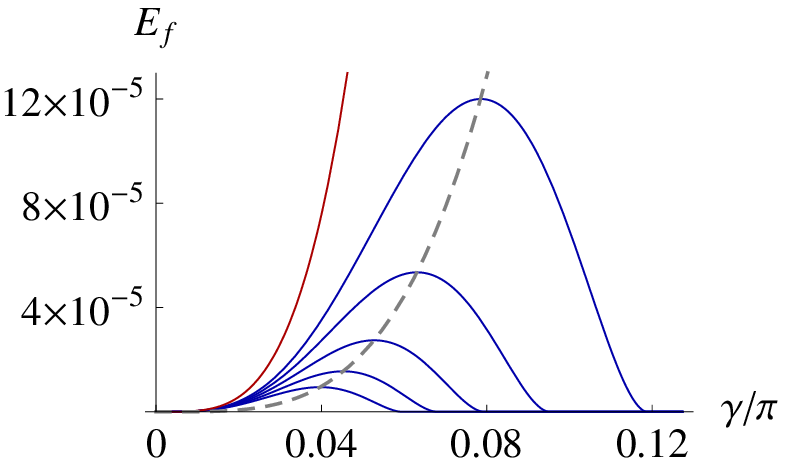,width=0.45\linewidth}}\hspace{5mm}
\caption{\label{fig:decoherence} The entanglement of formation $E_f$ of the two-atom state as a function of the phase-space angle $\gamma$ for $y\equiv\frac{2\lambda|\alpha|^3}{\Omega_1+\Omega_2}=0,0.15,0.4,0.7,1$ respectively (left) and $y=0,4,5,6,7,8$ (right) from top to bottom, where $\lambda$ is the cavity loss rate and $\Omega_{1,2}$ are the atom-field interactions. The dashed line in the right panel traces the maxima of $E_f$ over $\gamma$ for given values of $y$.}
\end{figure*}

The preceding results apparently contradict the correspondence principle by displaying entanglement, a purely quantum phenomenon, at the classical limit. However, a given value of entanglement is reached for a fixed $\gamma=\frac{\Omega_1t_1}{|\alpha|}=\frac{\Omega_2t_2}{|\alpha|}$, and therefore the required interaction time diverges when $|\alpha|\to\infty$. This is a demonstration of the singular nature of the semiclassical limit: for a fixed interaction time classical physics is recovered as $|\alpha|\to\infty$, but the classical and the long-time limits do not commute. A similar phenomenon was observed in \cite{lg}, where the classical limit and the limit of large squeezing do not commute.

Physically, longer interaction times imply a stronger effect of environment coupling, and it is this effect that guarantees the emergence of classical physics for large $|\alpha|$, both because of stronger dephasing and because of longer interaction times. We demonstrate this statement using the standard Markovian model of cavity loss, with Lindblad generator $a$ and rate $\lambda$, so that the master equation for the full system density matrix $\rho$ is \cite{zurek}
\begin{equation}\label{eq:master}
\textstyle\partial_t\rho=-\frac{i}{\hbar}[H,\rho]+\frac12\lambda(a^\dagger a\rho-2a\rho a^\dagger+\rho a^\dagger a)
\end{equation}
A superposition of distinct coherent states $c(\ket{\alpha}+\ket{\beta})$ experiences as a result of cavity loss, in addition to the overall decay with rate $\lambda$, a much faster dephasing with rate $\lambda|\alpha-\beta|^2$ that affects the coefficients of the coherence terms in the density matrix proportional to $\ket{\alpha}\bra{\beta}$ and its conjugate \cite{zurek}.

In order to analyze this process in conjunction with the wave packet dynamics we assume that $\lambda|\alpha|\ll \Omega_1,\Omega_2$ so that we can ignore the energy decay of the state. The full-system density matrix after the interaction with the first atom is then
\begin{align}
\rho_1&=\ket{1_+}\!\bra{1_+}+x_1\ket{1_+}\!\bra{1_-}+x_1^*\ket{1_-}\!\bra{1_+}
+\ket{1_-}\!\bra{1_-}
\end{align}
with 
$
x_k=e^{-\frac{\lambda|\alpha|^3}{\Omega_k}(\gamma+i(e^{-i\gamma}-1))}.
$
A similar dephasing affects the coefficient of the twelve coherence terms during the interaction with the second atom. 

The significance of dephasing for the generation of two-atom entanglement is the reduction of the coherence between the two wavepackets $\ket{2_{+-}}$ and $\ket{2_{-+}}$ that generates the entangled atomic state $\ket{c}$ when they collide, so that the term $\ket{2_{+-}}\bra{2_{-+}}$ and its conjugate in the full-state density matrix are multiplied by $x_1x_2$
and $(x_1x_2)^*$ (respectively). It follows that the term $\frac12 \ket{c}\bra{c}$ in $\rho_a$ is replaced by
\begin{align}
\rho_c&=\textstyle\frac14(\ket{c_A} \bra{c_A} +x_1x_2\ket{c_A}\bra{c_B}+(x_1x_2)^*\ket{c_B}\bra{c_A}
\nonumber\\&+ \ket{c_B} \bra{c_B})
\label{eq:od-dec}\end{align}
with 
\begin{equation*}
\ket{c_{A,B}}=\textstyle\frac12(\ket{g}_1\mp e^{ \frac{i}{2}\gamma_0}e^{\pm\frac{i}{2}\gamma}\ket{e}_1)\otimes(\ket{g}_2\pm e^{\frac{i}{2}\gamma_0}\ket{e}_2)
\end{equation*}
where upper (lower) signs correspond to $A$ ($B$), respectively, 
while the terms proportional to $\ket{l}\bra{l}$ and $\ket{r}\bra{r}$ are negligibly affected by decoherence. For $|x_1x_2|<1$ $\rho_c$ is the density matrix of a mixed state. Since $\ket{c_A}$ and $\ket{c_B}$ are product states, the degree of entanglement of $\rho_c$, and therefore $\rho_a$, is a decreasing function of the decoherence rate $\lambda$.

It follows from Eq.\ (\ref{eq:od-dec}) that the degree of entanglement depend on the decoherence rate only through the combination $y=\frac{2\lambda|\alpha|^{3}}{\Omega_1+\Omega_2}$. Fig.\ \ref{fig:decoherence} shows the entanglement of formation as a function of $y$ and $\gamma$. Evidently, the maximum achievable entanglement decreases with increasing $y$, and this maximum is achieved earlier. Detailed analysis shows that when $y$ is large the maximum of the entanglement is achieved at $\gamma=\frac1y$, and its value is proportional to $y^4\log y$. This phenomenon has the simple interpretation that a shorter effective interaction time allows less time for entanglement generation, but also less time for decoherence, and that this trade-off leads to a shorter optimal interaction time for stronger decoherence; for large $y$ the entanglement is completely destroyed when $\gamma=\frac{3}{2y}$. In this limit the interaction stops before the initial wave packet splits so that entanglement is generated only by weak quantum fluctuations, and this allows a power law rather than an exponential decay in the degree of entanglement---the entanglement of formation for the optimal interaction time is approximately $\frac19$ of the entanglement achieved in an {ideal} system for the same interaction time, see Fig.\ \ref{fig:decoherence} (right panel).

\section{Conclusions}
Our first conclusion is that two qubits \emph{can} be entangled solely by interaction mediated by a macroscopic field preparation, and the degree of entanglement tends to a positive value in the classical limit. The entanglement process proceeds through splitting of the field wave packet into four components each carrying a different atomic product state and a subsequent merging of two of the sub wave packets to form a entangled superposition of two atomic product states. Thus, although the initial field state is semiclassical with a large photon number and well-defined quadratures, the field necessarily evolves into highly non-classical states during the entanglement generation process. Nevertheless, semiclassical wave packet dynamics approximation for the propagation of the wave packet stays valid during the \emph{full} entanglement generation process, although in a multiple phase space---one phase space copy for each sub wave packet.

The splitting of the wave packet is a slow process where a macroscopic field state evolves by interaction with two qubits.
Naturally, it is not an efficient method of creating entangled pairs, as the interaction time required to obtain entanglement diverges in the classical limit. The correspondence principle is therefore nonetheless obeyed in the sense that classical physics, \textit{i.e.}\/ no entanglement, is obtained in the limit of large photon number for a \emph{fixed} interaction time.

The final two-atom state is a coherent superposition of two product states associated with the center wave packet, incoherently mixed with two additional products associated with the flank wave packets. Thus, the two-atom entanglement is determined by the overlap  of the evolving atomic states. In particular, wave packet splitting and atom-field entanglement are necessary but not sufficient for the generation of atom-atom entanglement.

A subtler issue is the fact that the two atoms are not maximally entangled by interaction with the semiclassical field mode. Although for short interaction periods the center wave packet has an almost Bell atomic state, the atom-atom entanglement is degraded almost completely by mixture with the flank wave packets, while for long interaction times the field and atoms decouple, the atomic state at this stage is separable. Unlike the vacuum field, therefore, the semiclassical field does not function as a quantum gate, and we conjecture that this is an example of a general principle.

We thank D Aharonov, I Arad, H Eisenberg, and N Katz, for helpful discussions. This work was supported by the ISF grant {1002/07}.


\begin{thebibliography}{99}
\bibitem{ens1} E Hagley et al., 
 Phys. Rev. Lett. \textbf{79}, 1 (1997).
\bibitem{eberly} JH Eberly, NB Narozhny, and JJ Sanchez-Mondragon, Phys. Rev. Lett.  \textbf{ 44}, 1323 (1980)
\bibitem{geaprl} J. Gea-Banacloche, Phys. Rev. Lett. \textbf{65}, 3385 (1990)
\bibitem{geapra} J. Gea-Banacloche, Phys. Rev. A \textbf{44}, 5913 (1990)
\bibitem{jc} O. Gat, Phys. Rev. A, \textbf{77}, 050102(R) (2008)
\bibitem{mondragon} SM Chumakov, A.B Klimov, and JJ Sanchez-Mondragon, Phys. Rev. A  \textbf{49}, 4972 (1994)
\bibitem{tcm}T Tessier, I Deutsch, A Delgado, I Fuentes-Guridi, Phys. Rev. A  \textbf{68}, 062316 (2003)
\bibitem{tcm2}  Jarvis et al.  New J. Phys. \textbf{11}, 103047 (2009)
\bibitem{nayak1} B Ghosh, AS Majumdar, and N Nayak, Int. J. Theor. Phys. Group Theory Nonlinear Opt. \textbf{ 13},  87-97 (2009)
\bibitem{nayak2}  B Ghosh, AS Majumdar, and N Nayak, Int. J. Quantum Inf. \textbf{5}, 169 (2007)
\bibitem{ens-dec} L. Davidovich, M. Brune, J. M. Raimond and S. Haroche, Phys. Rev. A, \textbf{53}, 1295 (1996)
\bibitem{bppaper} H.-P. Breuer, U. Dorner and F. Petruccione, Eur. Phys. J. D {\bf 14}, 377 (2001)
\bibitem{qd} D. Z. Rossatto, T. Werlang, E. I. Duzzioni, and C. J. Villas-Boas, \texttt{arXiv:1102.0073} (2011)
\bibitem{guo} S.B Zheng, G.C Guo, Phys. Rev. Lett. \textbf{85}, 2392 (2000)
\bibitem{ens2} S Osnaghi et al., 
Phys. Rev. Lett. \textbf{87}, 37902 (2001)
\bibitem{lj86} RG Littlejohn, Phys. Rep. \textbf{138}, 193 (1986).
\bibitem{wootters}W.K. Wootters, Phys. Rev. Lett. \textbf{80}, 2245 (1998)
\bibitem{hor}R.\ Horodecki, P.\ Horodecki, M.\ Horodecki, and K.\ Horodecki, Rev. Mod. Phys. \textbf{81}, 865 (2009) 
\bibitem{lg}A. Leshem and O. Gat, Phys. Rev. Lett. \textbf{103}, 070403 (2009)
\bibitem{zurek} W. H. Zurek, Rev. Mod. Phys. \textbf{75},  715 (2003).
\end{thebibliography}
\end{document}